# In-situ high resolution TEM observation of Aluminum solid-state diffusion in Germanium nanowires: fabricating sub- 10 nm Ge quantum dots

M.A. Luong[a], E. Robin[a], N. Pauc[b], P. Gentile[b], M. Sistani[c], A. Lugstein[c], M. Spies[d], B. Fernandez[d], M. I. den Hertog[d]*

a. Université Grenoble Alpes, CEA, INAC, MEM, Grenoble F-38000, France
b. Université Grenoble Alpes, CEA, INAC, PHELIQS/SINAPS, F-38000 Grenoble, France
c. Nanocenter Campus-Gußhaus, Institute of Solid State Electronics, Technische Universitat Wien, Gußhausstraße 25, Vienna 1040, Austria
d. Université Grenoble Alpes, CNRS, Institut NEEL UPR2940, 25 Avenue des Martyrs, Grenoble 38042, France

*«martien.den-hertog@neel.cnrs.fr»

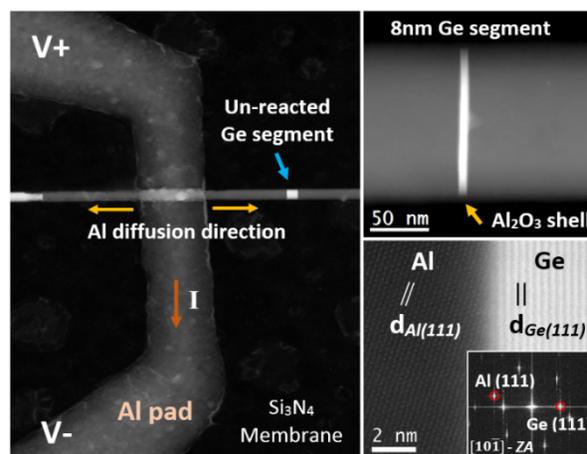

**ABSTRACT:** Aluminum- germanium nanowires (NWs) thermal activated solid state reaction is a promising system as very sharp and well defined one dimensional contacts can be created between a metal and a semiconductor, that can become a quantum dot if the size becomes sufficiently small. In the search for high performance devices without variability, it is of high interest to allow deterministic fabrication of nanowire quantum dots, avoiding sample variability and obtaining atomic scale precision on the fabricated dot size. In this paper, we present a reliable fabrication process to produce sub- 10 nm Ge quantum dots (QDs), using a combination of ex-situ thermal annealing via rapid thermal annealing (RTA) and in-situ Joule heating technique in a transmission electron microscope (TEM). First we present in-situ direct joule heating experiments showing how the heating electrode could be damaged due to the formation of Al crystals and voids at the vicinity of the metal/NW contact, likely related with electro-migration phenomena. We show that the contact quality can be preserved by including an additional ex-situ RTA step prior to the in-situ heating. The in-situ observations also show in real-time how the exchange reaction initiates simultaneously from several locations underneath the Al contact pad, and the Al crystal grows gradually inside the initial Ge NW with the growth interface along a Ge(111) lattice plane. Once the reaction front moves out from underneath the contact metal, two factors jeopardize an atomically accurate control of the Al/Ge reaction interface. We observed a local acceleration of the reaction interface due to the electron beam irradiation in the transmission electron microscope as well as the appearance of large jumps of the interface in unpassivated Ge wires while a smooth advancement of the reaction interface was observed in wires with an $Al_2O_3$ protecting shell on the surface. Carefully controlling all aspects of the exchange reaction, we demonstrate a fabrication process combining ex-situ and in-situ heating techniques to precisely control and produce axial Al/Ge/Al NW heterostructures with an ultra-short Ge segment down to 8 nanometers. Practically, the scaling down of Ge segment length is only limited by the microscope resolution.

I. Introduction

Over the past few years, nanowires[1–5] or nanotubes[6,7] as the building blocks for one dimensional (1D) materials have attracted enormous research interests for high performance devices[5,8] due to their notable properties of high surface area to volume ratio. The main challenge of nanoscale devices to daily applications is the formation of Schottky contacts between metallic and semiconducting interfaces causing a loss in electron transport[9,10]. To establish a low contact resistance between nanowires and metallic electrodes, it is of fundamental importance to understand and control the quality of the contacts. Recently, the formation of intermetallic Silicide or Germanide contacts via a thermally activated solid state reaction between the metal and Si or Ge NW has drawn significant attention because of its great advantages for fabricating short channel devices from bottom up grown NWs rather than complex and high-cost photolithography top-down approaches. With respect to the main stream of Si technologies, Ge can be preferable because of its unique properties of high carrier mobility and larger exciton Bohr radius enabling fabrication of high performance devices[11]. Numerous metals used as diffusion sources have been actively investigated such as Ni[5,12], Pt[5,13], Cu[11], Au[14] for high electrical conduction or Co[15], Fe[16], Mn[17] for magnetic applications. In this paper, Aluminum as a metallic source was chosen to elucidate the contact formation during the exchange reaction between the metallic contact electrode and Ge NW. It has been reported[18–21] that the intrusion of Al in Ge NWs will result in the formation of a monocrystalline Al core and multiple shells of Ge and $Al_2O_3$, respectively. In this paper we present real-time observations of the thermally induced solid state reaction of the Al/Ge binary system with the aim to deterministically fabricate Ge QD's with atomic size control between perfectly sharp metal contacts. To this end, we carefully study this exchange reaction in-situ from the nucleation stage of the propagation which starts at the NW surface underneath the contact electrode, to the end of the exchange process with atomically accurate control of the reaction interface position. We will show how the direct Joule heating, where a current is passed through a metal strip defined on the NW[22], can damage the heating electrodes, and it is therefore necessary to have an additional ex-situ RTA step prior to the in-situ heating to preserve the heating electrode. We also present the effect of the electron beam current density on the progress of the reaction interface and show the influence of an $Al_2O_3$ protecting shell around the NW on the kinetics of the exchange reaction. These effects should be optimized to have a better control of the diffusion process. Finally, we demonstrate the fabrication of a sub- 10 nm semiconductor quantum dot using the combination of ex-situ and in-situ heating methods, which could be a key for the future production of ultra-scaled devices.

Experimental

For the experiments, we used single-crystalline Ge NWs synthesized via the VLS process by the chemical vapor deposition method (CVD) with gold catalyst on (111) silicon substrate. $GeH_4$ gas was used as the precursor for the nanowire growth. The typical diameter of as-synthesized Ge nanowires varies around 100 – 150 nm with 8 μm length. The nanowires were lightly intentionally doped with Phosphorous. In our experiments, as grown Ge NWs were either being used directly to perform the metal contacts or the native $GeO_2$ shell formed due to the exposure of the NWs to the atmosphere was first removed by dipping in diluted hydriodic acid (HI acid, >57%, Sigma-Aldrich - MERCK) with deionized water (DI) with ratio 1:3 for 5 s and immediately coated with 5 nm of $Al_2O_3$ by atomic layer deposition (ALD) at 250 °C to protect them from oxidation and improve the surface quality. To apply the metal contacts, the NWs were first diluted in ethanol solution by sonication and dispersed on electron transparent 40 nm $Si_3N_4$ membranes by drop casting. The membrane fabrication has been described elsewhere[23,24]. SEM images of Ge NWs were taken for NW selection and designing the electrodes. Selected single or double Ge NWs were contacted by a pair of parallel Al rectangular bars so that the contacted NWs should be located near the center of each bar. Practically, for a good coverage of the NW surface, the optimized sizes of Al heating electrodes are about 650 nm in width and 5 um in length. The two ends of each Al bar were connected to the outer circuit by interconnecting Gold electrodes, patterned in a prior lithography step. The patterning process was defined using ebeam lithography with PMMA 4% photoresist. Before performing the metal deposition, the NWs with the protecting $Al_2O_3$ shell were dipped into buffered hydrofluoric acid - BOE 7:1 (HF : $NH^4F$ = 12.5 : 87.5%) for 10 s to completely remove the $Al_2O_3$ shell in the contact region and then put in diluted hydriodic acid (HI) for 5 sec to remove the $GeO_2$ shell. The samples were then cleaned by Ar plasma for 15 s before performing 200 nm Al deposition by sputtering (with the purity of 99.995% and in vacuum at a pressure lower than $10^{-6}$ Torr). Finally, the samples were lifted off in acetone solution overnight. The heating experiments were done by two different approaches: i.e., i) using a direct Joule heating method as described in the paper by Mongillo et al[22]. Particularly, a voltage difference ($V_{ap}$) was applied through the two ends of the Al metal strip and then slowly increased step by step of 0.025 V to increase the heating current ($I_h=V_{ap}/R_{strip}$). The resistance of the sputtered Al strips is around 130 Ω. The main advantage of this approach is performing the Al/Ge exchange reaction in a single step; however, the contact electrodes could easily damage at high applied voltage and current, before the exchange reaction actually started. ii) The samples were first heated using RTA in $N_2$ ambient at 300 °C during 20 s and cooled down to room temperature during 4 min. The RTA experiment was performed in a Jipelec™ JetFirst RTP Furnace[25]. RTA initiates the nucleation stage of the reaction in all contacts. Once Al has entered the Ge NW at the contact pad, the direct Joule heating can be performed without the need to apply high potentials/currents to the Al heating electrode, allowing propagation of the Al in the Ge NWs without damage to the Al contact. For temperature calibrated in-situ heating experiments, the samples were heated inside the TEM microscope using a commercial DENSsolution[26] six contact double tilt TEM holder connected with two 2401 Keithley sourcemeters providing the applied voltage. High angle annular dark field (HAADF) scanning TEM (STEM) was performed on a probe corrected FEI Titan Themis operating at 200 kV. The experiments were performed in STEM mode with a beam convergence angle of 20 mRad, the electron beam spot size 5 and gun lens 1.5.

## Results and discussion

### The nucleation of the exchange reaction and degradation of the Al heater by Joule heating

Consecutive heating via rapid thermal annealing is a common technique for performing a metal/semiconductor NW thermal exchange reaction. Fabrication of an ultra-short Si segment between PtSi contacts has been reported in literature[3] from Pt thermal diffusion in a Si NW. However, the reproducibility of such a diffusion process is typically not precise at atomic length scales, due to the fact that the exchange reaction does not start in all NWs at exactly the same moment, or to different reaction speeds depending either on NW diameter or contact quality[27]. Direct Joule heating, first reported in the work of Mongillo[22], is an interesting technique to combine heating and biasing experiments in a single sample geometry, and allows heating while observing the diffusion process in the electron microscope using and in-situ biasing TEM sample holder. However, one drawback of this technique is that the heating electrodes can damage easily to initiate the reaction, which will be demonstrated in the following experiments.

During the NW selection process for making the Al contact electrodes on the $Si_3N_4$ membranes, sometimes we obtained two NWs lying side-by-side on the membrane, that we decided to contact. This allows us to investigate the diffusion behavior in both NWs with slightly different diameters under exactly identical conditions. Figure 1 presents the HAADF STEM images of Aluminum incorporation into two Ge NWs having 5 nm $Al_2O_3$ shell. The video of this experiment is available as supporting information **M1**. The contrast in HAADF STEM is related both to the sample thickness and the atomic number of the elements present. Since Ge is heavier than Al, the brighter contrast corresponds to the initial Ge NW, and the darker contrast to the entering Al metal, as the NW dimension does not change much due to the exchange reaction. The diameters of the left and right NW are about 130 nm and 120 nm, respectively. The heating experiment was performed by passing a current from the left to the right of the Al metal strip (the electron flow is in the opposite direction) while monitoring the HAADF STEM image and contrast in the NWs. We slowly increased the applied voltage ($V_{ap}$) and observed the start of the exchange reaction at a defect, a small indentation at the surface in the middle of the left NW (indicated by a yellow arrow in figure 1a) when $V_{ap}$=1.7 V and $I_h$=3.5 mA. Then multiple nucleation points appeared at the NW surface. Clearly, even though the two NWs were contacted under identical conditions, the exchange reaction in the two NWs was different as the reaction of the right NW was a bit slower than the left NW (Figure 1b-f). The scenario of the exchange reaction could be: Ge atoms from the nanowire gradually dissolve into the Al reservoir by removing layer by layer of Ge(111) plane while Al atoms from the heating electrode diffuse into the NW to fill the empty space. This can be understood since the Ge(111) lattice spacing (0.32667 nm) is the largest interplanar distance of all crystal planes in Ge; its bonding energy is correspondingly the lowest and this plane is therefore most suitable for the exchange reaction to take place

(Figure 1c). It can be noted in **M1** that the reaction front (Al-Ge interface) stopped at the two edges of the Al contact pad and only started moving out when Ge atoms underneath the Al pad had fully dissolved in both NWs. This diffusion behavior results in a symmetric Al propagation length on both sides of the Al contact electrode. The annealing process also created some voids in the Al metal contact in the vicinity of the contacted NW, shown in figure 1e-f.

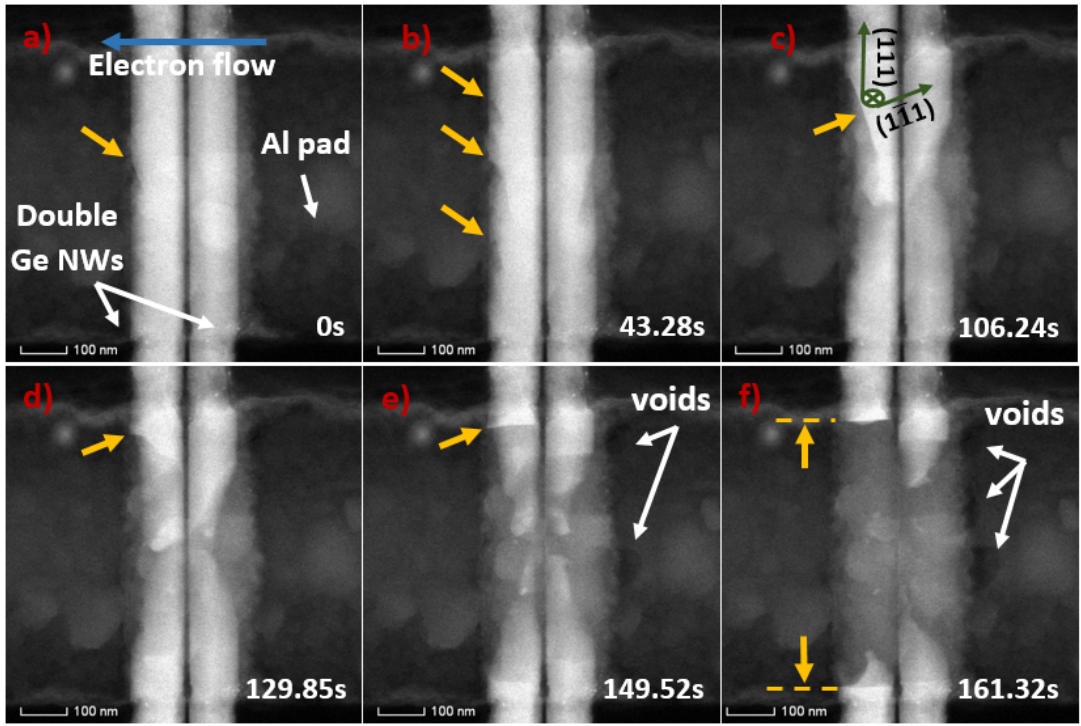

**Figure 1.** HAADF-STEM image sequences of thermally assisted diffusion of Al into double Ge nanowires based on Joule heating method. A heating current is passed through the Al heating electrode from the left to the right, as indicated by the blue arrow (the flow of the electrons is reversed). **a)** The propagation nucleated from a defect at the surface of a Ge NW (yellow arrow). **b)** Multiple reacting points appeared simultaneously when raising the applied voltage, accelerating the exchange process. **c-e)** The reaction front moved along the direction of (111) planes until reaching the edges of the Al contact pad. **f)** The exchange reaction rate of the right NW is a bit slower than the left NW. The exchange reaction also induced voids on the right side of the Ge NWs. The scale bar is 100 nm. See the supporting information **M1**.

In another experiment, presented in SI **M2**, we observed the formation of Al crystals on the heating strip, and some remaining Ge islands present below the contact electrode. Figure 2 shows the HAADF images of Al incorporation in a 162 nm diameter Ge NW with a 5 nm $Al_2O_3$ shell. The exchange reaction started at $V_{ap}$=0.725 V and $I_h$= 3.2 mA. As can be seen in figure 2a, even after the reaction front had moved out from underneath the Al metal contact pad, some Ge islands (indicated by white arrows) remain present in the NW underneath the contact pad. Besides, when the heating temperature was increased, a large Al crystal formed on the Al contact (Figure 2b-f). The formation of Al crystals on the heating pads was typically observed in most of the heating electrodes (images not shown). The redistribution of Al atoms by the replacement with Ge atoms within the NW and the 'uphill' diffusion into the large Al crystal caused a local exhaustion of Al atoms from the contact pad. Consequently a cracking point appeared on the right side close to the NW surface (Figure 2f). We have also performed ex-situ heating experiments on similar devices using rapid thermal annealing, and no damage was found at the Al contact. Hence, the degradation of the Al heating electrode in Joule heating process is likely due to the electro-migration effect of the heating current. The result is in good agreement with reports[28–30] where the current will result in the formation of voids on the negative electrode and Al islands tend to form on the positive side.

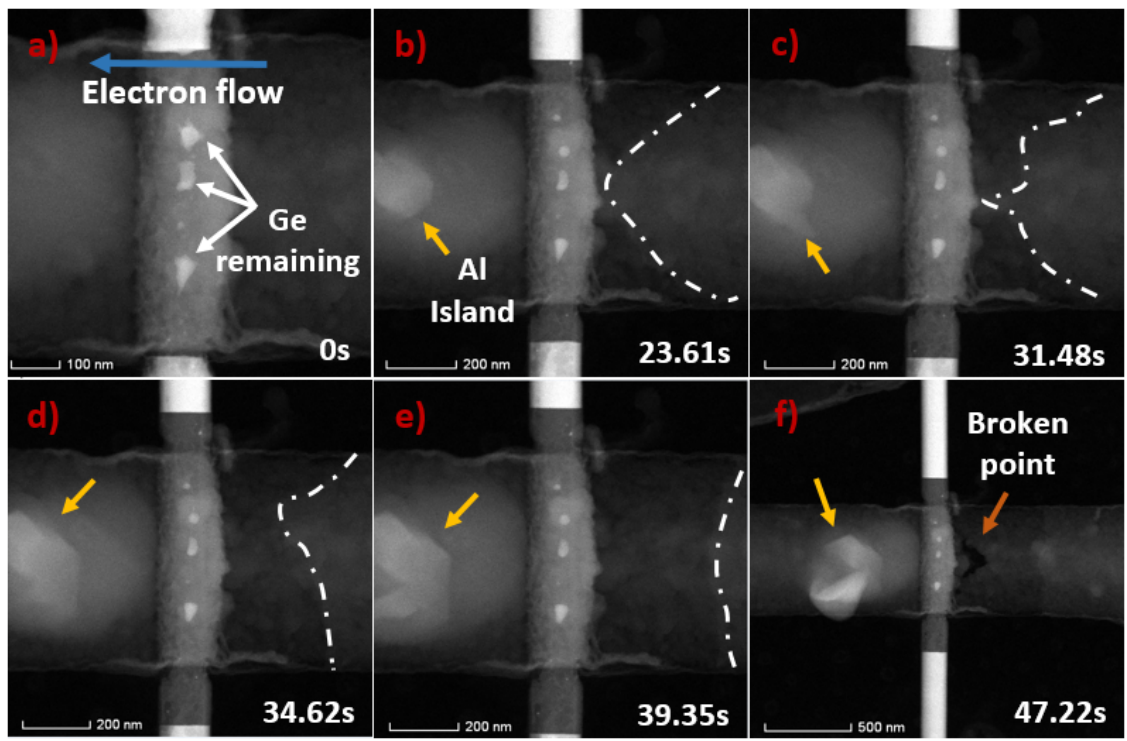

**Figure 2.** HAADF-STEM images of Al intruding in 130 nm Ge NW with 5 nm $Al_2O_3$ shell by Joule heating method. **a)** Some remaining Ge islands are present underneath the Al contact pad when the reacting front moved out. **b-e)** The diffusion of Al atoms into the Ge NW and the formation of a large Al crystal (yellow arrows) lead to a local exhaustion of Al atoms on the contact pad. **f)** A cracking point appeared on the right side of the NW in the Al contact, creating an open contact and blocking the heating current. The orange arrow indicates a broken point in the Al electrode. The scale bar varies in each image. See the supporting information **M2**.

From these experiments we saw that the initiation of the exchange reaction can occur at multiple nucleation points below the Al contact, and that the reaction interface moves out from this contact only once all the Ge underneath the contact has been consumed. Using the Al metal strip as a direct Joule heater to initiate the solid state exchange reaction can also damage the Al contact, most likely due to electro-migration phenomena and the relatively high current that is needed to start the exchange reaction. We will show that this can be mitigated using first an ex-situ RTA of the specimen to initiate the exchange reaction and then continue the Al propagation process in a direct Joule heating experiment. This approach helps to preserve the contact quality and reduces the heating current during the in-situ annealing process. Using a combination of both ex-situ and in-situ annealing we could perform experiments in a more reliable way and were able to study the reaction interface in high detail.

During the in-situ heating experiments, we have seen an influence of the electron beam on the exchange reaction. Moreover, we monitored the reaction kinetics and observed an influence of the NW surface quality on the reaction kinetics and the height of the nucleation step at the reaction interface. Hence for an atomically precise control of the diffusion process, especially for producing sub- 10 nm semiconductor segments, it is important to master these effects during the sample preparation and TEM observation.

### Effect of electron beam on the reaction interface propagation rate

It is well known in TEM investigations that the electron beam can damage the specimen if it stays on the specimen for a certain time. This can be due to the high energy of the electron beam that causes a displacement of atoms from the lattice above the displacement threshold of the investigated materials and therefore artificially perturbs the vacancy concentration from equilibrium[31]. Since we are performing heating experiments, another potential explanation can be additional heating due to the electron beam. In our experiments, presented in supporting information **M3** we have seen a clear influence of the electron beam on the diffusion behavior at the Al/Ge interface, in the case that the electron beam scanned only a small region of the reaction interface and then zoomed out to lower magnification for observation. The experiments were performed in STEM mode with a probe current of $I_b$=96 pA. During STEM HAADF serial imaging, the electron beam was scanned over a 512 x 512 raster with a per-pixel dwell time of 2 µs on the lateral direction from the left to the right and from the top to the bottom in a square field of view of 193.28 nm x 193.28 nm. The electron dose rate (e. $nm^{-2}$. $s^{-1}$) was calculated by dividing the probe current by the area of the raster, which is about $1.61 \times 10^4$ (e. $nm^{-2}$. $s^{-1}$). Figure 3a-f shows the influence of the electron beam on the Al/Ge binary exchange reaction in the 133 nm Ge NW when it was interacting with the specimen. As indicated by the orange arrow in figure 3a, the Al migration front had advanced faster at the location where the beam had scanned a small area in previous frames. When the beam then scans a larger area including the entire reaction interface (Figure 3b-f), the difference between the middle front and the two edges becomes smaller, and finally the interface becomes straight again. Similar observations have been reported in the work of Fauske et al.[32] where the thermal diffusion of Au in GaAs nanowires was studied. Similar to their observations, the beam induced increased replacement rate was limited to the regions near the electron beam exposed area, and did not extend across the entire nanowire cross-section. Therefore, once a step has nucleated, it does not necessarily advance over the entire reaction interface, indicating that the nucleation event is not the rate limiting step of this reaction.

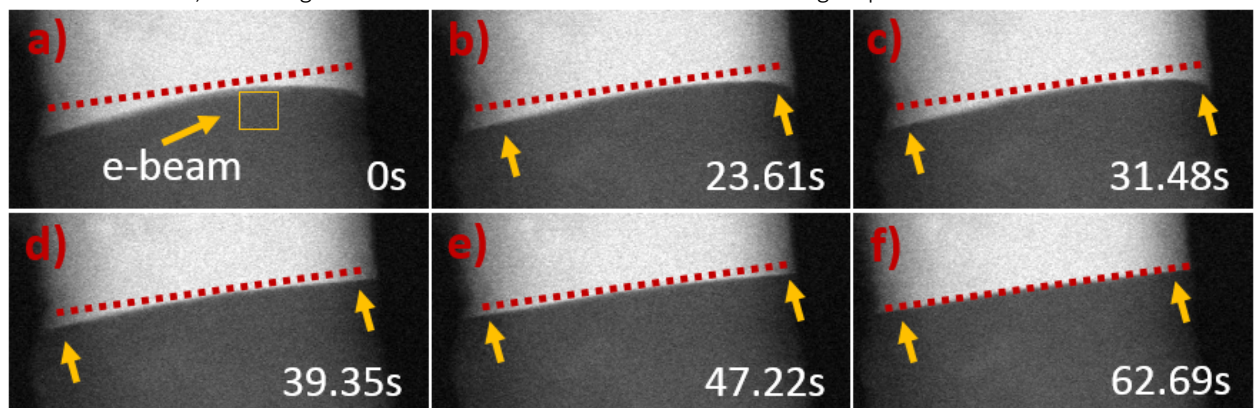

**Figure 3**. Sequences of HAADF-STEM images were taken while the specimen was being irradiated by the electron beam. **a)** Image obtained after zooming out from the indicated region. **b-f)** After decreasing the magnification, the reaction interface is progressively restored to a flat interface. The yellow arrows indicate the regions far from the more irradiated square indicated in (**a**). See Supporting Information **M3**.

Radial and axial propagation of the Al/Ge interface – effect of $Al_2O_3$ shell

The diffusion rate could also be affected by other factors; for instance, in the study of Holmberg et al.[33], they have reported the growth rate dependence on the distance from the metal source. Yaish et al.[27], have showed in the Ni/Si system that the quality of the nanowire surface determined by the exposure time to ambient air before the metal deposition step could change the growth rate from the square root to the linear time dependence. In the present paper, we present clear evidence of an effect of the NW surface quality on the reaction kinetics and demonstrate that the surface quality can be improved by the presence of a protecting shell. Figure 4a-f represent HAADF-STEM images of Al migration in a 133 nm Ge NW without the presence of an $Al_2O_3$ shell. These NWs were exposed to the atmosphere so that they have been oxidized forming a $GeO_2$ shell (a 3D reconstruction of the $GeO_2$ shell has been reported in the work of El Hajraoui[34]). Before starting the heating experiment, the NW was oriented such that the NW Ge(111) growth plane is parallel to the electron beam to allow better observation of the nucleation phenomena at the Ge(111) reaction interface. The Joule heating was started and series of HAADF images were acquired during the exchange reaction. The diffusion behavior in this NW is presented in the supporting information **M4**. As shown in figure 4a, the interface of Al-Ge was initially very sharp. After 0.787 s, from the edge of the NW, there was a jump of the Al reacting front of 6.5 nm in the axial direction, corresponding to a replacement of 20 layers of Ge(111) atomic planes. Then the Al/Ge reacting front runs in the lateral direction from the left to right and finishes a cycle after 7.87 s. The diffusion speed in the lateral direction could be estimated to be about 16.9 nm/s. Figure 4g shows the plot of Al protruding length **L** as a function of time in the axial direction of the propagated NW without having the $Al_2O_3$ shell. The graph demonstrates local variations in propagation speed of the Al/Ge reaction front, due to the nucleation and subsequent ledge flow of steps of a few nm in height on the reaction interface. As can be seen from the plot, the average propagation rate of the reaction interface is about 0.64 nm/s. It is worth noting that the growth rate in the axial direction is 26 times smaller than in the lateral direction. This result is quite reasonable since it took some time to nucleate a new step on the reaction interface. From literature[35], it has been known that the Ge surface is very sensitive to atmospheric ambience. Probably, the NW surface was strongly oxidized, and this may be related to increased surface roughness and/or the formation of surface defects at the Ge/GeO$_2$ interface. Hence the trapping and de-trapping of the reaction interface at these defects could result in the observed stepwise growth. In contrast, investigating the Al/Ge exchange reaction in a Ge NW with the presence of a 5 nm Al$_2$O$_3$ shell, a smoother linear time dependence of the Al diffusion length is observed and plotted in figure 4h. The reaction interface speed is about 0.143 nm/s. It is worth noting that the diffusion speed in this case is lower than in the example of a NW without the protecting shell (0.64 nm/s). However, we cannot interpret the kinetics between the two NWs since they propagated at different temperatures (different heating current) and the Joule heating experiments are not temperature calibrated. These results may indicate that in the latter case the quality of the NW surface was better controlled by the protection of the Al$_2$O$_3$ layer, hence the exchange reaction took place uniformly along the NW axial direction without the strong fluctuation of the diffusion rate due to surface defects. A smooth advancement of the interface is paramount for controlling the length of the semiconductor region with atomic scale spatial resolution.

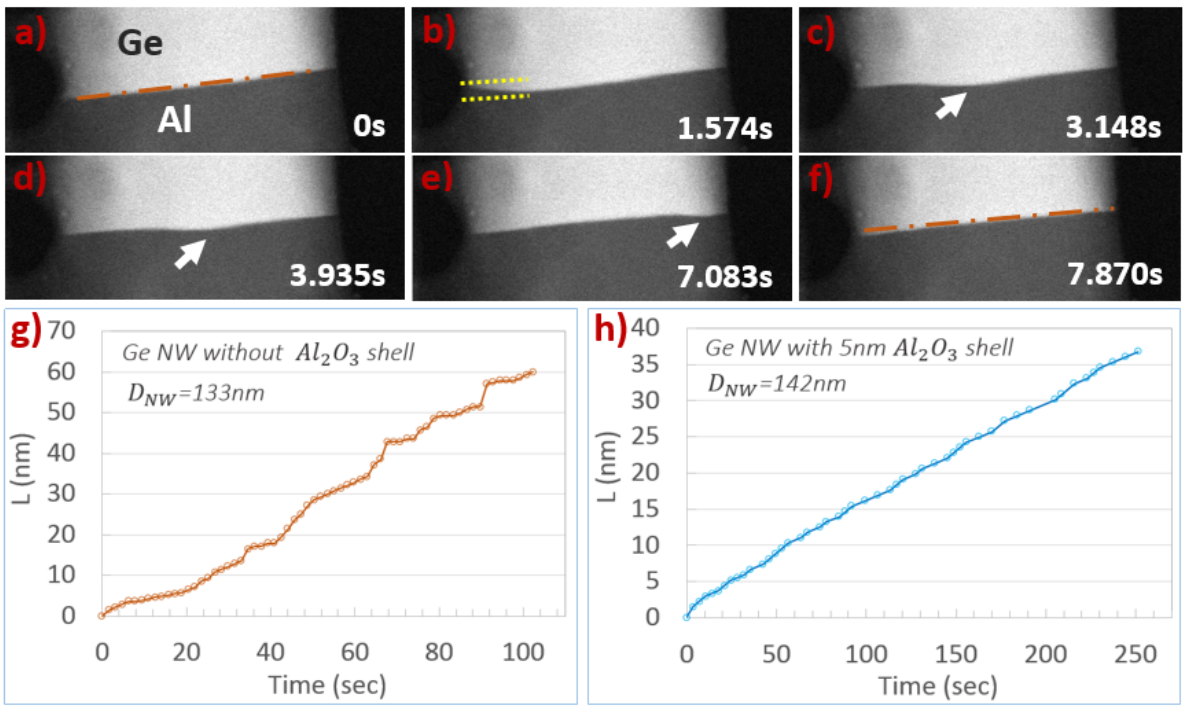

**Figure 4. a-f)** Series of HAADF-STEM images of Al migration in a Ge NW showing the nucleation of a step at the Al/Ge reaction interface and its propagation in the lateral direction from the left to the right. The diameter of the NW is about 133 nm. The yellow dash lines present the step height. The white arrows indicate the positions of the step moving along the reaction interface from the left to the right. **Figure 4(g-h)** shows the plot of Al converted length *L* as a function of time in the axial direction of the NW without and with the protection of a 5 nm Al$_2$O$_3$ shell, respectively. The movie of the diffusion presenting the experiment of Fig. 4(a-g) is available for viewing in Supporting Information **M4**. The experiment presented in Fig. 4h is available in the SI **M5**.

It should be noted that the current study focuses on relatively large diameter NWs (120≤d≤162). Previously we have observed a smooth propagation behavior in much thinner NWs (15≤d≤50) at low temperatures (250 – 330 °C) where the length of the propagated region as a function of time follows a square root behavior, indicating the reaction is limited by diffusion[34]. In addition using an ex-situ reaction at a heating temperature of 350 °C, it was observed that no propagation occurs in NW diameters above 150 nm[18]. These combined observations potentially indicate that the reaction rate is limited by diffusion for small NW diameters, and is limited by an interface (either the metal reservoir-NW interface or the reaction interface) for larger diameter NWs.

Combining ex-situ and In-situ heating for producing an ultra-small Ge QD

Using the direct Joule heating technique, it is quite difficult to generate the diffusion gently without damaging the heating electrodes. Therefore we first annealed the sample by RTA to initiate the exchange reaction between Ge and Al. Since the whole sample was heated at the same temperature, no damage was found at the contact pads after the intrusion of Al into the Ge NW. Figure 5a-c represents the HAADF-STEM images of Al incorporation in a 142 nm diameter Ge NW with 5 nm $Al_2O_3$ shell, to obtain a smooth advancement of the reaction interface. The sample was heated at 300 °C for 20 s in $N_2$ ambience, giving 750 nm propagation length from each side (figure 5a). We then performed the direct Joule heating sequentially from each side to carefully reduce the unreacted Ge segment length. The movie for the production of a small Ge segment is presented in supporting information M5. When the Ge segment reached sub- 10 nm scale, it is important to lower the applied heating voltage to decrease the diffusion rate until achieving the desired segment length. Figure 5b illustrates the formation of an ultra- small 8 nm Ge segment after a RTA treatment followed by an in-situ Joule heating process. The segment length of this Ge disk was intentionally stopped at 8 nm for further investigations. Indeed, it is also possible to perform a complete Al-Ge exchange for the fabrication of a monocrystalline Al NW[21]. Figure 5c shows an HR-HAADF-STEM image of the interface between the reacted and unreacted part with the corresponding FFT in the inserted figure. The interface appears quite sharp when oriented on the [10$\bar{1}$] zone axis of Al. There is a rotation of 18 degrees between the Al(111) atomic planes and Ge(111) planes on the growth direction as can be observed in the FFT shown in the inset of figure 5c.

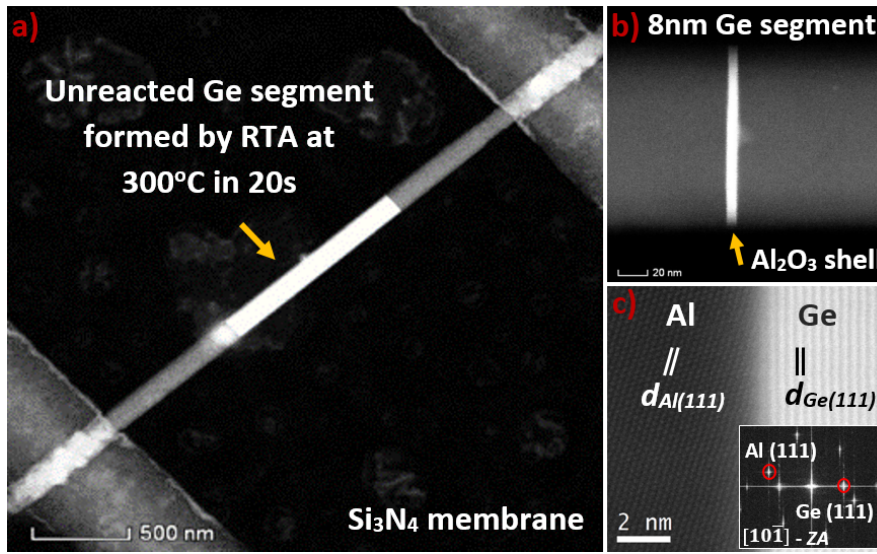

**Figure 5. a)** HAADF-STEM image of an axial Al/Ge/Al NW hetero-structure formed after the RTA treatment at 300 °C for 20 s. The NW diameter is about 142 nm and covered by 5 nm $Al_2O_3$ shell. **b)** An ultra- short 8nm Ge segment was created by combining RTA and in-situ Joule heating method. **c)** HR-STEM image with the corresponding FFT at the Al/Ge interface showing the 18 degree rotation between Al(111) and Ge(111) planes when the specimen is oriented on the [10$\bar{1}$] direction of observation on the Al part. The interface appears very sharp. See Supporting Information M5.

CONCLUSION

In summary, the diffusion behavior of the Al-Ge thermal exchange reaction was fully described from the earliest stage to the end of the exchange process using real-time observations, providing a method to deterministically create a semiconductor region with atomic scale precision. It is clear that the diffusion rate can depend on numerous factors such as Al-Ge contact quality, heating temperature and the presence of the electron beam. Importantly, we have demonstrated the influence of the NW surface quality on the diffusion behavior of Al in a Ge NW. Without the presence of a protecting layer, the Al protruding length shows stepwise growth dependence as a function of time while a linear time dependence will take place when the NW is covered by a 5 nm $Al_2O_3$ shell in these relatively large diameter NWs (120≤d≤162). Finally, combining in-situ and ex-situ heating processes, we have successfully controlled and synthesized an ultrashort Ge segment of 8 nm. From literature, this is the smallest Ge segment length that has been obtained in the Al/Ge binary system, and among the smallest segment in other semiconductor NW metal systems. These atomic-scale observations are extremely useful to understand the reaction behaviors of metals with respect to Ge NWs or other semiconductors and observe potential differences with respect to bulk materials.

## ABBREVIATIONS:

**FETs**: field effect transistors; **CVD**: chemical vapor deposition; **VLS**: vapor-liquid-solid; **NWs**: nanowires; **ALD**: atomic layer deposition; **STEM**: scanning transmission electron microscope; **RTA**: rapid thermal annealing; **HAADF**: high-angle annular dark-field imaging.

## ASSOCIATED CONTENT
* Supporting Information

The Supporting Information is available free of charge on the ACS

Publications website at DOI: ………

SI movie **M1** shows the incorporation of Aluminum into double Ge NWs (130 nm and 120 nm in diameter, from the left to the right, respectively) having 5 nm $Al_2O_3$ shell. SI movie **M2** shows HAADF-STEM images of Al intruding in 130 nm Ge NW with 5 nm $Al_2O_3$ shell by Joule heating method. SI movie **M3** presents the influence of the electron beam to the diffusion behavior of the Al/Ge binary system. SI movie **M4** demonstrates the stepwise diffusion rate of Al/Ge thermal exchange in the 133 nm Ge NW without having the protecting shell. SI movie **M5** shows the production of an ultra- small Ge segment from the combination of ex-situ and in-situ heating techniques.

* Competing interests:

The authors declare that they have no competing interests.

## AUTHOR INFORMATION
Corresponding Authors
*E-mail: martien.den-hertog@neel.cnrs.fr

## ACKNOWLEDGMENTS
The authors would like to thank Dr. Laurent CAGNON for help on the ALD technique and Dr. Stephane Auffret for help with the Al sputtering. We acknowledge support from the Laboratoire d'excellence LANEF in Grenoble (ANR-10-LABX-51-01). We benefitted from the access to the Nano characterization platform (PFNC) in CEA Minatec Grenoble and Nano-fab from institute NEEL, Grenoble. We acknowledge funding from the ANR in the T-ERC project e-See. This project has received funding from the European Research Council (ERC) under the European Union's Horizon 2020 research and innovation programme (grant agreement N° 758385) for the e-See project. The authors gratefully acknowledge financial support by the Austrian Science Fund (FWF): project No.: P28175-N27.